# DYNAMICALLY AND THERMALLY NONEQUILIBRIUM FLUCTUATION-ELECTROMAGNETIC INTERACTIONS: RECENT RESULTS AND TRENDS

GEORGE DEDKOV[+] and ARTHUR KYASOV[++]

*Nanoscale Physics Group, Kabardino-Balkarian State University n. a. Kh. M. Berbekov,*

*Nalchik, 360004 Russia*

[+]gv_dedkov@mail.ru, [++]aa_kyasov@mail.ru

Our recent theoretical developments related to the nonrelativistic and relativistic fluctuation-electromagnetic interactions of bodies with different temperatures moving translationally and (or) rotationally relative to each other are briefly summarized. Three basic geometrical configurations and physical systems are discussed: "a small particle and a thick plate" (i); "a small particle in a radiative vacuum background" (ii) and "two thick plates in relative motion" (iii) – classical Casimir-Lifshitz configuration with allowance relative motion and different temperatures of plates. For configuration 3, it is shown that the theory of friction and heat exchange by Levin-Polevoi-Rytov proves to be quite adequate contrary to the settled point of view of many authors.

## 1. Introduction

Currently, interest in the issues of dynamically and thermally nonequilibrium fluctuation-electromagnetic interactions (FEI) is stimulated by the development of microelectromechanical (MEMS) and microoptomechanical (MOMS) systems,[1-3] studies of rapidly rotating particles in atomic traps,[4–7] and astrophysical applications.[8–11] Moreover, for more than 35 years, there is a steady interest in studying quantum or Casimir friction between the bodies in relative motion (see Refs. 12–14 and many references therein). Recently, stable rotation of an optically trapped dielectric particle of 100 nm diameter at rotation frequencies exceeding 1 GHz was observed in Refs. 3 and 4. Optically levitated particles have the potential to reach extremely high rotation speeds[4,5] necessary to study unexplored types of FEI and vacuum friction effects. In this paper we give a brief description of our recent results[14-23] related to FEI of bodies in several configurations at thermal and dynamical disequilibrium.

## 2. Rotating Particle Near a Surface

### 2.1 *Nonretarded approximation*

Details of the solution to this problem are given in Refs. 14 and 21. In what follows, all the quantities (if not mentioned) are defined in the reference coordinate systems of a thick plate or a vacuum background (both are at rest). We considered the general case when the rotation axis of the particle is directed at an arbitrary angle $\theta$ relative to the plane of a thick plate. The particle is assumed to be a point fluctuating dipole characterized by the dipole polarizability $\alpha(\omega)$ and local temperature $T_1$, located at a distance $z_0$ from the plate, and the plate material is characterized by the local dielectric permittivity $\varepsilon(\omega)$ and temperature $T_2$. Hereinafter the temperatures are expressed in the units of energy. The main characteristics of FEI in this configuration are the force **F**, torque **M**, and the rate of particle heating, $dQ/dt$. It is worth noting that **F** in this case has two nonzero components, one of which is the familiar Casimir-Polder force $F_z$, normal to the plate (including the angular velocity dependence). Another (lateral) component, $F_y$, appears as a net result of particle rotation. The corresponding expressions have the form

$$F_z = -\frac{3\hbar}{32\pi z_0^4} \int_{-\infty}^{+\infty} d\omega \left[ (2-\cos^2\theta) A_1(\omega,T_1,T_2) + (2+\cos^2\theta) A_2(\omega,T_1,T_2) \right] \quad (1)$$

$$A_1(\omega,T_1,T_2) = \Delta'(\omega)\alpha''(\omega)\coth\left(\frac{\hbar\omega}{2T_1}\right) + \Delta''(\omega)\alpha'(\omega)\coth\left(\frac{\hbar\omega}{2T_2}\right) \quad (2)$$

$$A_2(\omega,T_1,T_2) = \Delta'(\omega)\alpha''(\omega_+)\coth\left(\frac{\hbar\omega_+}{2T_1}\right) + \Delta''(\omega)\alpha'(\omega_+)\coth\left(\frac{\hbar\omega}{2T_2}\right) \tag{3}$$

$$F_y = \frac{3\hbar\cos\theta}{16\pi z_0^4}\int_{-\infty}^{+\infty}d\omega\,\Delta''(\omega)\alpha''(\omega_+)\left[\coth\left(\frac{\hbar\omega_+}{2T_1}\right) - \coth\left(\frac{\hbar\omega}{2T_2}\right)\right] \tag{4}$$

In (1)–(4), $\omega_+ = \omega + \Omega$, $\Delta(\omega) = (\varepsilon(\omega)-1)/(\varepsilon(\omega)+1)$, and the quantities with one and two dashes denote the relevant real and imaginary components. From (7) we see that lateral force is zeroed when $\theta = \pi/2$, i.e. when the particle spin is oriented normal to the surface. In its turn, the particle heating rate is

$$dQ/dt = -\frac{\hbar}{16\pi z_0^3}\int_{-\infty}^{+\infty}d\omega\,\omega\Delta''(\omega)\left[(2-\cos^2\theta)B(\omega,\omega) + (2+\cos^2\theta)B(\omega_+,\omega)\right] \tag{5}$$

$$B(x,y) = \coth\left(\frac{\hbar x}{2T_1}\right) - \coth\left(\frac{\hbar y}{2T_2}\right) \tag{6}$$

Expressions for the projections of torque **M**, as well as the consideration of particle dynamics were presented in Refs. 14 and 21. An important result is that angular velocity $\Omega$ and orientation angle $\theta$ at an arbitrary instant of time are interrelated via the expression

$$\Omega = \Omega_0\frac{\sin\theta\tan^2\theta_0}{\sin\theta_0\tan^2\theta} \tag{7}$$

where $\Omega_0$ and $\theta_0$ are the initial values at $t=0$. Equation (7) implies that $\theta \to \pm\pi/2$ at the particle braking stage depending on the sign of $\theta_0$. Thus, for any initial orientation, the spin of particle tends to orient perpendicular to the surface, but the states $\theta \to \pm\pi/2$ are reached only at the full stop being asymptotically stable. The state with the rotation axis parallel to the surface is unstable and with any small deviation from this state the modulus of angle $\theta$ would increase. The change in the azimuthal angle $\varphi$ (in the plane parallel to the plate) does not affect $\Omega$ and $\theta$, causing the angular momentum precession. In the case of a plate without dielectric losses ($\varepsilon'' = 0$), we obtain $d\varphi/dt = const$, $\theta = \theta_0 = const$, $\Omega = \Omega_0 = const$. Such a situation, however, is typical only for a nonretarded interaction with the plate. Accounting for retardation[14] shows that the particle rotation is damped due to electromagnetic radiation.

## 2.2. *Nonthermal radiation near a transparent dielectric plate*

As shown above, the angular velocity vector of a dielectric particle rotating in the near field of a surface tends to become normal to it. Therefore, it is this configuration of most interest. When we have a moving (rotating) particle and a stationary fluctuating electromagnetic field of another (transparent) body, the radiation intensity of the particle can be expressed through the Joule dissipation integral inside the particle:

$$I = \oint_\sigma S_n\cdot d\sigma = -\int_\Omega \langle\mathbf{j}\cdot\mathbf{E}\rangle d^3r \equiv I_1 - I_2 \tag{8}$$

where $S_n$ is the normal projection of the Pointing vector $\mathbf{S} = (c/4\pi)\langle\mathbf{E}\times\mathbf{H}\rangle$ to the wave surface $\sigma$ surrounding the particle, $\Omega$ (do not mix with angular velocity) is the particle volume, $I_1, I_2$ are the intensities of emitted and absorbed radiation ($I_2 = 0$ in the case of nonthermal radiation). Equation (8) is a consequence of the energy conservation law when the electromagnetic field energy is constant. For a moving and rotating particle, Eq. (8) can be rewritten in the form[14-18]

$$I = I_1 - I_2 = -(dQ/dt + \mathbf{F}\cdot\mathbf{V}) \tag{9}$$

Where **F** is the force, $dQ/dt = \gamma^{-2}(dQ''/dt'' + M_n'\Omega)$, $dQ''/dt''$ is the net heating rate of the particle in its rest frame, $M_n'$ is the torque specified in the coordinate system moving with velocity $\mathbf{V}$, $\gamma = (1-\beta^2)^{-1/2}$, $\beta = V/c$. For a transparent dielectric, at $V=0, T_1 = T_2 = 0$ [14]

$$I = -\frac{\hbar}{\pi c^3}\int_0^\Omega d\omega\,\omega^4 \sum_{s=e,m}\alpha_s''(\Omega-\omega)\psi_s(n,\omega z_0/c) \tag{10}$$

In (10), $n$ is the refraction index of the dielectric plate material (ferroelectric, for example), and functions

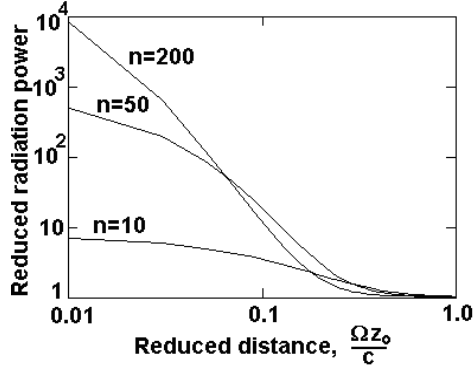

Fig. 1. Reduced total radiation power $(I + I^{(vac)})/I^{(vac)}$ of a gold particle near the surface of a dielectric plate with refractive index $n$, $I^{(vac)}$ is the intensity of radiation in vacuum (see Sect. 3).

$\psi_s(n, \omega z_0/c)$ ($s=e,m$) are given in Ref. 14. Typically, for $n >> 1$ we have $|\psi_m(n,x)| >> |\psi_e(n,x)| >> 1$, and therefore, the intensity of radiation (Fig. 1) is greatly amplified compared to the radiation of a particle rotating in vacuum (see Sect. 3). In plotting Fig.1, we used the low-frequency Drude approximation of the particle polarizability, $\alpha''(\omega) = 3R^3\omega/4\pi\sigma_0$ ($R$ and $\sigma_0$ are the particle radius and static conductivity). As follows from Fig. 1, the intensity of radiation drustically increases with $n$ at distances satisfying weakly depends on the distance to the plate if $\Omega z_0/c << 1$ ($z_0 << 1 cm$ for realistic values $\Omega < 10^{10} s^{-1}$).

## 3. Translational-Rotational Motion and Radiation of a Particle in a Vacuum Background

In line with the argumentation given in Sect. 3.1 (Eqs. (8), (9)), we consider a neutral spherical particle with radius $R$ and temperature $T_1$ moving with an arbitrary velocity $\mathbf{V}$ and angular velocity $\Omega$ ($\Omega R/c << 1$) in radiation background with temperature $T_2$. The spin of particle is assumed to have the angle $\theta$ relative to $\mathbf{V}$ and lies in the plane $xz$ of the Cartesian reference system of the background radiation, velocity $\mathbf{V}$ coincides with the $x$ axis. The resultant intensity of radiation has the form[19]:

$$I = I_1 - I_2 = \frac{\hbar\gamma}{4\pi c^3} \int_{-\infty}^{+\infty} d\omega \omega^4 \int_{-1}^{1} dx\, F(\omega, x) \tag{11}$$

$$F(\omega, x) \equiv \alpha''(\omega_\beta^-) f_1(x, \beta, \theta) B(\omega_\beta^-, \omega) + \alpha''(\omega_\beta^+) f_1(x, \beta, \theta) B(\omega_\beta^+, \omega) \tag{12}$$

$$f_1(x, \beta, \theta) \equiv (1-\beta^2)(1-x^2)\cos^2\theta + 0.5\left((1+\beta^2)(1+x^2) + 4\beta x\right)\sin^2\theta \tag{13}$$

$$f_2(x, \beta, \theta) = (1-\beta^2)(1-x^2)\sin^2\theta + 0.5\left((1+\beta^2)(1+x^2) + 4\beta x\right)(1+\cos^2\theta) \tag{14}$$

where $\alpha'' = \alpha_e'' + \alpha_m''$ is the sum of the electric and magnetic polarizabilities, $\beta = V/c$, $\gamma = (1-\beta^2)^{-1/2}$, $\omega_\beta^- = \gamma\omega(1+\beta x)$, $\omega_\beta^+ = \omega_\beta^- + \Omega$. In Eq. (11), intensities $I_1, I_2$ are associated with the terms depending on $T_1$ and $T_2$. Expression (11) describes both the nonthermal radiation intensity and the thermal contributions. The expression for the nonthermal radiation intensity follows from (17) using the limiting transitions $T_1 \to 0, T_2 \to 0$. The resultant expression has the form[19]

$$I = -\frac{\hbar\gamma}{2\pi c^3}\int_{-1}^{1} dx \cdot f_2(x,\beta,\theta) \int_0^{\Omega\gamma^{-1}(1+\beta x)^{-1}} d\omega\omega^4 \alpha''(\omega_\beta^-) = \frac{4\hbar}{3\pi c^3}\int_0^\Omega d\xi \xi^4 \alpha''(\Omega - \xi) \tag{15}$$

From (15) we see that the integral intensity of nonthermal radiation does not depend on linear velocity $V$, but the spectral-angular intensity distribution essentially depends on the relativistic factor and mutual angular orientation of the linear and angular velocity vectors. Figure 2 shows the normalized spectral intensity distributions $I(\omega)/I_0$ calculated in the case of gold particles with the polarizability $\alpha(\omega)$ used in Sect. 2.2. The normalization factor is $I_0 = (3\hbar\Omega^2/4\pi^2\sigma_0)(\Omega R/c)$.

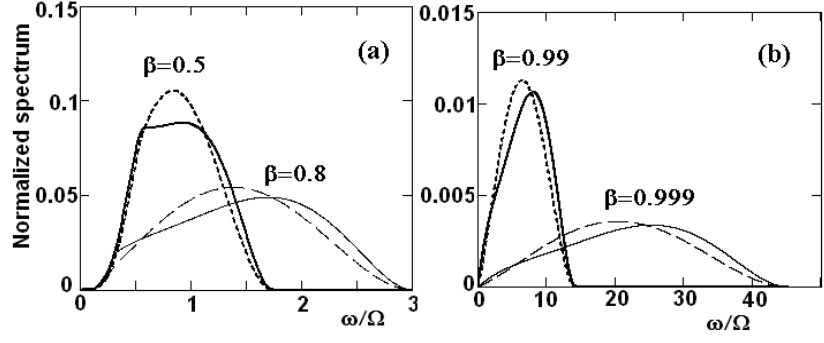

Fig. 2 Normalized frequency spectrum of nonthermal radiation[16]. Solid and dashed lines correspond to parallel and perpendicular spin orientations relative to the linear velocity vector.

In this case, the tangential force $F_x$, the heating rate $dQ/dt$ and the torque projections $M_x, M_z$ are given by simple relations[19]

$$F_x = -\beta I/c, \; dQ/dt = -I/\gamma^2, \; M_x = -4\hbar\cos\theta/(3\pi c^3\gamma)J(\Omega), \; M_z = \gamma M_x \tan\theta \qquad (16)$$

where $J(\Omega)$ is the same integral as in (15) when replacing $\xi^4 \to \xi^3$. Since the particle mass $m$ changes due to radiation as $d(mc^2) = \gamma dQ/dt$ (Refs. 14 and 17), the dynamics equation $(d/dt)(\gamma mV) = F_x$ takes the form $\gamma^3 m dV/dt = F_x - \gamma V dm/dt$, with the right-hand part to be zero due to (16). Therefore, $V = const$ and radiation is due to the particle rotation in accordance with Eq. (15), which does not depend on $V$.

The analysis of rotational dynamics shows that angular velocity $\Omega$ decreases with time, whereas $\theta = const$. The possible spin precession around vector **V** has no effect on the intensity of radiation. The rotation energy is transformed into radiation, but not entirely. Some part of energy causes the heating of particle[19], which leads to the instability of thermal state $T_1 = T_2 = 0$. Due to this, after some time, the radiation intensity will contain a thermal component[19].

Thermal configuration $T_1 \neq 0, T_2 \neq 0$ is more complicated, but, as in the "cold" case, acceleration $d\beta/dt$ does not depend on the particle temperature and angular velocity. The radiation intensity depends on $\Omega$ and $\beta$, while the spin tends to orient perpendicularly to the linear velocity. It is interesting, that the particle may quickly reach the state of thermal quasiequilibrium depending on $T_2, \Omega$ and $V$, and the maximum power radiated in this state is proportional to $\gamma^2$.[15,16,19] It turns out that the time of thermal relaxation ($\tau_Q$) is much less than the times of stopping for both rotational ($\tau_\Omega$) and translational ($\tau_\beta$) dynamics: $\tau_Q \ll \tau_\Omega \ll \tau_\beta$.[19] Therefore, when calculating the time dependence of the ratio $T_1/T_2$, one may fix parameters $\Omega$ and $\beta$, whereas when calculating the time dependences of $\Omega$ and $\theta$, one may fix $T_1/T_2$ and $\beta$.

## 4. Friction and Heat Exchange Between Parallel Thick Plates in Relative Motion

This problem has a very long history (for a review see Refs. 12–14). However, strange as it may seem to be, the correct solution to it was obtained already in 1990 by Polevoi[24] within the fluctuation-electrodynamic approach developed in Ref. 25 for the problem of heat exchange in the same configuration (hereafter Levin-Polevoi-Rytov (LPR) theory). Unfortunately, unlike the seminal paper by Lifshitz,[26] who calculated the attractive force between thick plates at rest, results[24,25] were not recognized until now (see, for example, Refs. 12, 13 and 27–29). Meanwhile, as we have shown recently,[22,23] the basic expressions obtained later by many authors for the friction force and heat exchange in the discussed configuration exactly follow from the LPR theory. This overlook is likely due to the fact that in Ref. 24, the friction force is zeroed in the limit $c \to \infty$ due to the use of a simplified impedance approximation for metal plates.

Assuming that the lower plate is at rest and the upper one is moving in the $x$ direction with velocity $V$ parallel to the plates, in the most interesting case $V/c \ll 1$ with allowance for retardation, we transformed the basic expressions in Ref. 24 for the friction force (per unit surface area of plates) to the form[22,23]

$$F_x = \frac{\hbar}{4\pi^3}\int_0^\infty d\omega \int d^2k \, k_x \left[ \operatorname{Im}\left(\frac{q_1}{\varepsilon_1}\right) \operatorname{Im}\left(\frac{\tilde{q}_2}{\tilde{\varepsilon}_2}\right) \frac{|q|^2}{|Q_\varepsilon|^2} + (\varepsilon \to \mu) \right] \cdot B(\omega, \tilde{\omega}) \qquad (17)$$

$$Q_\varepsilon = (q + q_1/\varepsilon_1)(q + \tilde{q}_2/\tilde{\varepsilon}_2)\exp(qa) - (q - q_1/\varepsilon_1)(q - \tilde{q}_2/\tilde{\varepsilon}_2)\exp(-qa) \tag{18}$$

where $q_i = (k^2 - (\omega/c)^2 \varepsilon_i \mu_i)^{1/2}$, $i=1,2$; $q = (k^2 - (\omega/c)^2)^{1/2}$; $\tilde{\omega} = \omega - k_x V$. In (17), (18), $a$ is the gap width between the plates, $\varepsilon_{1,2}, \mu_{1,2}$ correspond to the dielectric permittivity and magnetic permeability of the upper (1) and lower (2) plates, and quantities with tilde depending on $\omega$ are taken at $\omega = \tilde{\omega}$; the temperature factor $B(\omega, \tilde{\omega})$ is given by Eq. (6). The heat flux per unit surface area of plate 1 is given by the same expression (17) when replacing $k_x \to -\omega$ in the integrand. In the first-order velocity approximation, at $T_1 = T_2 = T, k_x V \ll T/\hbar$, Eq. (17) reduces to[22,23]

$$F_x = \frac{\hbar V}{2\pi^2} \int_0^\infty d\omega \frac{dn(\omega)}{d\omega} \int_0^\infty dk k^3 \left[ \operatorname{Im}\left(\frac{q_1}{\varepsilon_1}\right) \operatorname{Im}\left(\frac{q_2}{\varepsilon_2}\right) \frac{|q|^2}{|Q_\varepsilon|^2} + (\varepsilon \to \mu) \right] \tag{19}$$

where $n(\omega) = [\exp(\hbar\omega/T) - 1]^{-1}$. Unlike (17), formula (20) does not contain variables with tilde. An important feature of Eqs. (17) and (20) is that $F_x$ is expressed through the dielectric permittivities and magnetic permeabilities and do not contain the reflection factors. Expression (29) can be easily transformed to the form[22,23] containing the reflection factors $\Delta_{i\varepsilon} = (\varepsilon_i q - q_i)/(\varepsilon_i q + q_i)$, $\Delta_{i\mu} = (\mu_i q - q_i)/(\mu_i q + q_i)$:

$$F_x = (\hbar/4\pi^3) \int_0^\infty d\omega \int_{k>\omega/c} d^2 k k_x \exp(-2qa) \operatorname{Im}\Delta_{1\varepsilon} \operatorname{Im}\tilde{\Delta}_{2\varepsilon} |D_\varepsilon|^{-2} B(\omega, \tilde{\omega}) + $$
$$+ (\hbar/16\pi^3) \int_0^\infty d\omega \int_{k\leq\omega/c} d^2 k k_x (1 - |\Delta_{1\varepsilon}|^2)(1 - |\tilde{\Delta}_{2\varepsilon}|^2) |D_\varepsilon|^{-2} B(\omega, \tilde{\omega}) + (\varepsilon \to \mu) \tag{20}$$

where $|D_\varepsilon| = |1 - \Delta_{1\varepsilon}\tilde{\Delta}_{2\varepsilon}\exp(-2qa)|$. Formula (20) completely coincides with the known results in this configuration.[12,13,29] For example, at $T_1 = T_2 = 0$, we retrieve from (20) the force of quantum friction[29]. We recalculated numerically the friction force between two Au plates using (19) and the Drude approximation for $\varepsilon(\omega)$ with $T$-dependent relaxation time of electrons (according to the Blokh-Gruneisen law). The absolute values of the friction force proved to be $10^7$ times higher than in Ref. 24 and, which is more intriguing –the friction force drastically increases at $T < 50K$.[22,23]

## 5. Promising Experiments

In this section, we feature in short possible experimental situations in which friction forces and the associated FEI effects could be observed, namely: (1) angular deflection and energy losses of accelerated ($V \sim 10^6 m/s$) neutral (or with a small charge) atomic, molecular or cluster beams at grazing incidence to the surface; (2) measuring the Casimir-Lifshitz friction forces in an AFM scheme when using probing metallic spheres with a diameter of 10-100 μm at low-temperature conditions (about 50 K or less); (3) damping and radiation of rotating particles near a dielectric surface with large refraction index *n* in the frequency range $\omega \sim \Omega$; (4) excess low-frequency thermal and nonthermal radiation from cosmic dust particles (rotating and/or moving translationally) and giant gas-dust clouds with internal dynamics. The dust particles are typically charged, but their electromagnetic radiation due to the presence of charge has another spectrum as compared to that caused by FEI, and therefore, both radiation spectra can be separated.

## 6. Conclusion

In dynamically and (or) thermally nonequilibrium systems, many interesting new phenomena arise, such as quantum or van-der-Waals-Casimir friction of particles in translational and (or) rotational motion, as well as specific thermal and nonthermal radiation. It turns out that dynamics, thermal evolution and electromagnetic radiation in such systems are strongly interrelated. We believe that experimental studying of these effects is very promising.